\documentclass[12pt]{article}

\usepackage{amssymb}
\usepackage{amsmath}
\usepackage{amscd}
\usepackage{latexsym}
\usepackage{url}

\usepackage{cite}

\usepackage{graphicx}
\usepackage{subfigure}

\newcommand{\be}{\begin{equation}}
\newcommand{\ee}{\end{equation}}

\newcommand{\dlt}{\delta}
\newcommand{\prt}{\partial}
\newcommand{\br}{{\bf r}}
\newcommand{\bk}{{\bf k}}

\newcommand{\bt}{\beta}

\newcommand{\ep}{\varepsilon}

\newcommand{\ra}{\rightarrow}
\newcommand{\sgm}{\sigma}

\newcommand{\gm}{\gamma}
\newcommand{\om}{\omega}

\newcommand{\dgr}{\dagger}
\newcommand{\lbd}{\lambda}
\newcommand{\Lbd}{\Lambda}

\newcommand{\rgl}{\rangle}
\newcommand{\lgl}{\langle}

\begin{document}

\begin{center}

{\Large{\bf Ground state of homogeneous Bose gas of hard spheres} \\ [5mm]

V.I. Yukalov$^{1}$ and E.P. Yukalova$^{2}$ } \\ [3mm]

{\it
$^1$Bogolubov Laboratory of Theoretical Physics, \\
Joint Institute for Nuclear Research, Dubna 141980, Russia \\ [3mm]

$^2$Laboratory of Information Technologies, \\
Joint Institute for Nuclear Research, Dubna 141980, Russia }

\vskip 2mm

\end{center}

\vskip 2mm

\begin{abstract}

The ground state of a homogeneous Bose gas of hard spheres is treated in 
self-consistent mean-field theory. It is shown that this approach provides 
an accurate description of the ground state of a Bose-Einstein condensed 
gas for arbitrarily strong interactions. The results are in good agreement 
with Monte Carlo numerical calculations. Since all other mean-field
approximations are valid only for very small gas parameters, the present 
self-consistent theory is a unique mean-field approach allowing for an accurate
description of Bose systems at arbitrary values of the gas parameter.  

\end{abstract}

\vskip 5mm


\newpage

\section{Introduction}

The quantum hard-sphere model serves as a reference or as an initial approximation
for quantum systems with more complicated interaction potentials because this model
is characterized only by a single interaction parameter composed of the system 
density and the sphere diameter. The interest to the hard-sphere Bose systems, 
initiated by the works of Bogolubov [1], Lee, Huang, and Yang [2-4], Wu [5], and 
others has been connected with the attempt to give a reasonable description for a 
quantum fluid with more realistic potentials, especially for liquid helium. By 
extensive numerical simulations, Kalos, Levesque, and Verlet [6] proved that the 
hard-sphere reference fluid is able to provide good description even for liquid 
helium, whose atoms interact through the Lennard-Jones potential. They showed that 
the attractive forces change the liquid structure only a little [6].        

The model characterizing the interactions in Bose systems by a single gas parameter 
has become intensively employed for low-temperature Bose gases, where at small values 
of the gas parameter the system properties are shown to be universal, being almost
independent on the particular shapes of interaction potentials [7]. Bose systems,
whose atomic interactions are characterized by a gas parameter, have been extensively
studied by Monte Carlo numerical calculations for both trapped [8-11] and homogeneous
gases [7,12-14].     

It would, certainly, be good to have a theory of a mean-field type, which could 
provide more or less simple formulas for treating Bose systems with finite gas 
parameters. However, there is a wide-spread consensus that there exist no theoretical 
description, based on a mean-field approximation, that could give reasonably accurate 
results outside of the region of very small gas parameters, where the Bogolubov 
approximation is valid. Actually, the Bogolubov approximation is often identified 
with the mean-field theory [7,9,12]. 

The absence for Bose-condensed systems of a mean-field approximation, that could give
at low temperatures a reasonable description for finite or large interactions, seems
rather strange, since for many other systems such mean-field approximations do exist. 
For example, many magnetic materials, defined by the Heisenberg or Ising models,
at low temperatures can be reasonably well described by the mean-field approximation.
Of course, a mean-field approximation can fail in the critical region or for reduced 
dimensions, but in three dimensions at very low temperatures, close to zero, such 
approximations do catch the main properties of magnetic materials [15,16].

In the present paper, we show that the low-temperature Bose systems are not outcasts
enjoying no accurate mean-field theory, but there exists a mean-field approach 
providing a correct description of such systems for arbitrarily large gas parameters 
and yielding the results in close agreement with numerical Monte Carlo calculations.

\section{Representative ensemble}

Our consideration is based on the self-consistent approach to Bose-condensed systems
[17-20], employing representative ensembles [21,22]. This approach guarantees the 
self-consistency of all thermodynamic relations, the validity of conservation laws,
and a gapless spectrum of collective excitations.  

The energy Hamiltonian for a Bose system of hard spheres is written in the standard 
form
\be
\label{1}
 \hat H = \int \hat\psi(\br) \left ( -\; \frac{\nabla^2}{2m} \right ) 
\hat \psi(\br) \; d\br \; + \; \frac{1}{2} \; \Phi_0 \int \hat\psi^\dgr(\br) 
\hat\psi^\dgr(\br) \hat \psi(\br) \hat \psi(\br) \; d\br \; ,
\ee
with the interaction strength 
\be
\label{2}
\Phi_0 \equiv 4\pi \;\frac{a_s}{m}
\ee
characterized by scattering length $a_s$ and atomic mass $m$. The field operators 
satisfy the Bose commutation relations. Generally, the operators depend on time 
which, for brevity, is not shown explicitly. Here and in what follows, the Planck 
and Boltzmann constants are set to one. 

Note that we take the interaction potential in the form of a local pseudopotential, 
which is admissible when the interaction radius is much shorter than mean 
interatomic distance. Strictly speaking, the scattering length represents the 
hard-sphere diameter only when the scattering length $a_s$ is essentially shorter 
than the interatomic spacing $a$. In that case, as is known [2-6,23], the results
for the local pseudopotential coincide with those for the hard-sphere system. The use
of the local pseudopotential for the finite values of the ratio $a_s/a$ can be 
justified by the following reasons. First of all, this ratio for a liquid cannot be 
larger than about $0.6$, since after this the liquid freezes [13]. More important 
is that the approximations we employ are based on the possibility of extrapolating
the results obtained for small parameters to the large values of these parameters.
Thus, the self-consistent mean-field approximation [18-20], we use, can be shown to 
be equivalent to a variational procedure with respect to atomic correlations, which
makes it possible to extend the results from the region of weak interactions to 
that of strong interactions. The self-similar approximation allows us to extrapolate 
the expressions, derived in the limit of small coupling parameters, to the region 
of large parameters, as has been demonstrated for a number of quantum models [24,25].
These methods guarantee that the results obtained for the small ratio $a_s/a$, where
$a_s$ well represents the hard-sphere diameter, provide us good approximations
for the finite values of this ratio.    
                 
The necessary and sufficient condition for the occurrence of Bose-Einstein 
condensation is the spontaneous breaking of global gauge symmetry [26,27]. The 
symmetry breaking can be explicitly realized by means of the Bogolubov shift [28,29]
for the field operator
\be
\label{3}
  \hat \psi(\br) = \eta(\br) + \psi_1(\br) \;  ,
\ee
where $\eta({\bf r})$ is the condensate wave function and $\psi_1(\bf r)$ is the field
operator of uncondensed atoms. It is worth stressing that the Bogolubov shift (3)
is not an approximation, but an exact canonical transformation [30].    

To avoid double counting, the condensate function and the field operator of 
uncondensed atoms are assumed to be orthogonal to each other,
\be
\label{4}
 \int \eta^*(\br) \psi_1(\br) \; d\br = 0 \;  .
\ee
The operator of uncondensed atoms on average is zero,
\be
\label{5}
 \lgl \psi_1(\br) \rgl = 0 \;  ,
\ee
so that the condensate function plays the role of an order parameter
\be
\label{6}
 \eta(\br) =  \lgl \hat\psi(\br) \rgl  \; .
\ee

By this definition, the condensate function and the field operator of uncondensed 
atoms are treated as separate variables [28,29], normalized, respectively, to the 
number of condensed atoms
\be
\label{7}
N_0 = \int | \; \eta(\br) \; |^2 \; d\br
\ee
and to the number of uncondensed atoms 
\be
\label{8}
 N_1 = \lgl \hat N_1 \rgl \; ,
\ee
where the operator of uncondensed atoms is
$$
\hat N_1 \equiv \int \psi^\dgr_1(\br) \psi_1(\br) \; d\br \;  ,
$$
and the total number of atoms in the system is $N = N_0 + N_1$.  

The evolution equations for the variables are obtained [17,18,22] by the 
extremization of the effective action, under conditions (4) to (8), which yields 
the equation for the condensate function
\be
\label{9}
i\; \frac{\prt}{\prt t} \; \eta(\br,t) = 
\left \lgl \frac{\dlt H}{\dlt \eta^*(\br,t)} \right \rgl
\ee
and the equation for the operator of uncondensed atoms
\be
\label{10}
i\; \frac{\prt}{\prt t} \; \psi_1(\br,t) = 
 \frac{\dlt H}{\dlt \psi_1^\dgr(\br,t)} \;  ,
\ee
with the grand Hamiltonian
\be
\label{11}
 H = \hat H - \mu_0 N_0 - \mu_1 \hat N_1 - \hat\Lbd \;  ,
\ee
in which
\be
\label{12}
 \hat\Lbd = \int \left [ 
\lbd(\br) \psi_1^\dgr(\br) + \lbd^*(\br) \psi_1(\br) 
\right ] \; d\br \;  .
\ee
The Lagrange multipliers $\mu_0$ and $\mu_1$ guarantee the validity of the 
normalization conditions (7) and (8), while the Lagrange multipliers 
$\lambda(\bf r)$ guarantee the conservation condition (5). These evolution 
equations are proved [31] to be identical to the Heisenberg equations of motion.

The system statistical operator in equilibrium is defined by minimizing the 
information functional [31,32] uniquely representing the system with the given 
restrictions. This results in the statistical operator
\be
\label{13}
 \hat\rho = \frac{1}{Z} \; e^{-\bt H} \qquad 
\left ( Z \equiv {\rm Tr} e^{-\bt H} \right ) \; ,
\ee
with the same grand Hamiltonian (11) and $\beta \equiv 1/T$ being the inverse 
temperature. 

For a system of $N$ atoms in volume $V$, the average density
\be
\label{14}
\rho \equiv \frac{N}{V}  = \rho_0 + \rho_1
\ee
is the sum of the densities of condensed and uncondensed atoms, respectively,
\be
\label{15}
 \rho_0 \equiv \frac{N_0}{V} \; , \qquad \rho_1 \equiv \frac{N_1}{V} \; .
\ee

For a homogeneous system, $\eta(\bf r) = \sqrt{\rho_0}$. The terms, containing 
the operators of uncondensed atoms, are treated in the Hartree-Fock-Bogolubov
approximation. The details of this self-consistent mean-field approach for Bose 
systems have been thoroughly exposed in Refs. [18-20,22,31], so that here we omit 
the intermediate calculations, passing to the final results. For the density of 
uncondensed atoms, we find 
\be
\label{16}
\rho_1 = \int \left [ \frac{\om_k}{2\ep_k} \; 
\coth \left ( \frac{\ep_k}{2T} \right ) -\; \frac{1}{2} \right ] \;
\frac{d\bk}{(2\pi)^3} \;   ,
\ee
where the notation
\be
\label{17}
\om_k \equiv \frac{k^2}{2m} + mc^2
\ee
is used, and the expression
\be
\label{18}
\ep_k = \sqrt{ (ck)^2 + \left ( \frac{k^2}{2m}\right )^2 }
\ee
represents the spectrum of collective excitations. The sound velocity $c$ is 
defined by the equation
\be
\label{19}
 mc^2 =  ( \rho_0 + \sgm_1 ) \Phi_0 \;  .
\ee
The anomalous average
\be
\label{20}
\sgm_1 = - \int \frac{mc^2}{2\ep_k}
\coth \left ( \frac{\ep_k}{2T} \right ) \; \frac{d\bk}{(2\pi)^3} 
\ee
describes the density $\vert \sigma_1 \vert$ of pair-correlated atoms [31].

\section{Zero temperature}

To consider the ground state, we set temperature to zero. Then the density of 
uncondensed atoms (16) becomes
\be
\label{21}
 \rho_1 =  \frac{(mc)^3}{3\pi^2}  \qquad ( T = 0 ) \; ,
\ee
while for the anomalous average, we have
\be
\label{22}
 \sgm_1 = - mc^2 \int  \frac{1}{2\ep_k} \; \frac{d\bk}{(2\pi)^3} \; .
\ee

This integral (22) for the anomalous average is divergent. This is why the often used
practice is to omit the anomalous average at all, just setting $\sigma_1$ to zero. This,
however, is principally wrong, since the nonzero anomalous average is the manifestation
of the broken gauge symmetry, in the same way as the nonzero condensate fraction. 
Omitting the former would require to neglect the latter, hence, would prohibit the 
condensate existence. It is straightforward to show that neglecting the anomalous 
average makes the system with Bose-Einstein condensate unstable [17,22,31,33]. 

The integral (22) can be regularized by invoking one of the known regularization 
procedures, all of which are actually equivalent to the dimensional regularization [34].    
Such a regularization is known to be asymptotically exact in the limit of weak 
interactions. Therefore, regularizing the integral in Eq. (22), one has to keep 
in mind the limit $\Phi_0 \ra 0$ in the spectrum (18), which can be taken into account
by replacing there $c$ with $c_{eff}$, such that
\be
\label{23}
 c_{eff} \simeq c_B \qquad ( \Phi_0 \ra 0 ) \;  ,
\ee
where 
\be
\label{24}
c_B \equiv \sqrt{ \frac{\rho}{m} \; \Phi_0 } 
\ee
is the asymptotic value of the sound velocity for $\Phi_0 \ra 0$, that is, the
Bogolubov sound velocity [1,28,29]. This yields
\be
\label{25}
 \int  \frac{1}{2\ep_k} \; \frac {d\bk}{(2\pi)^3} = 
- \; \frac{m^2}{\pi^2} \; c_{eff}  \;  .
\ee
Thus, the anomalous average (22) can be reduced to the form
\be
\label{26}
\sgm_1 = \frac{m^3c^2}{\pi^2} \; c_{eff} \qquad (\Phi_0 \ra 0 ) 
\ee
that is asymptotically exact in the limit of weak interactions [34]. 

Since we are interested in describing finite values of atomic interactions, the 
next step would be an analytic continuation of form (26) to finite $\Phi_0$. 
Before defining this procedure, let us pass to dimensionless quantities that will 
also be more convenient for numerical calculations. 

Let us define the fractions of condensed and uncondensed atoms, respectively,
\be
\label{27}
 n_0 \equiv \frac{\rho_0}{\rho} \; , \qquad
 n_1 \equiv \frac{\rho_1}{\rho} \; ,
\ee
and the dimensionless anomalous average
\be
\label{28}
 \sgm = \frac{\sgm_1}{\rho} \;  .
\ee
And let us introduce the dimensionless sound velocity
\be
\label{29}
 s \equiv \frac{mc}{\rho^{1/3}} \;  .
\ee

As a dimensionless strength of atomic interactions, it is natural to use the 
{\it gas parameter}
\be
\label{30}
 \gm \equiv \rho^{1/3} a_s \;  ,
\ee
which is of order of the ratio $a_s/a$. 

It is worth emphasizing that this parameter is natural, since it describes the 
ratio of the effective potential energy of an atom to its kinetic energy. Really,
potential energy per atom is proportional to $\rho a_s / m$, while kinetic energy 
is of order $\rho^{2/3} / m$. The ratio of the former to the latter gives exactly
the gas parameter (30).   

In the dimensionless units, the fraction of uncondensed atoms reads as
\be
\label{31}
 n_1 = \frac{s^3}{3\pi^2} \;  .
\ee
Equation (19) for the sound velocity transforms into 
\be
\label{32}
 s^2 = s_B^2 ( n_0 + \sgm) \;  ,
\ee
with the dimensionless Bogolubov velocity
\be
\label{33}
s_B \equiv \frac{mc_B}{\rho^{1/3}} = \sqrt{4\pi\gm} \;   .
\ee
The equation for the anomalous average (26) reduces to  
\be
\label{34}
 \sgm =\frac{s^2}{\pi^2} \; s_{eff} \qquad (\gm \ra 0 ) \; ,
\ee
where $s_{eff} = m c_{eff} / \rho^{1/3}$. 

The problem in extending the weak-interaction formula (34) to finite interactions
is the necessity of defining an analytic continuation from asymptotically small 
$\gamma \ra 0$ to the finite values of $\gamma$. Such an analytic continuation
seems to be not uniquely defined. For instance, if we set $s_{eff} = s_B$ in 
Eq. (34), we come back to a Bogolubov-type approximation that can be accurate
for small gas parameters $\gamma < 0.1$. Setting $s_{eff} = s$, we get the 
approximation of Ref. [35], valid for $\gamma < 0.2$. 

In order to extend the validity of approximations to larger values of $\gamma$, 
it is useful to keep in mind that, as has been stressed above, the nonzero 
anomalous average requires a nonzero condensate fraction, as far as both of them
arise due to the global gauge symmetry breaking occurring under Bose-Einstein 
condensation [26,27]. On the contrary, the zero condensate fraction implies the 
zero anomalous average, which writes as the condition
\be
\label{35}
 \sgm \ra 0 \qquad ( n_0 \ra 0 ) \;  .
\ee
The mentioned approximations $s_{eff} = s_B$ and $s_{eff} = s$ do not satisfy
condition (35), which explains why they do not allow for extending expression (34) 
to the values of the gas parameter larger than $\gamma < 0.2$.         

An approximation, satisfying condition (35), can be obtained by defining $s_{eff}$ 
from Eq. (32) by setting $\sigma$ to zero in the right-hand side of this equation, 
which gives $s_{eff} = \sqrt{4 \pi\gm n_0}$. This approximation was employed in 
Ref. [19], which allowed for the extension of the accurate results to $\gamma < 0.4$,
as compared with the Monte Carlo calculations [7,13].   

Now we propose a better justified procedure for analytically extending the 
anomalous average to higher values of the gas parameter. For this purpose, we rewrite
Eqs. (32) and (34) in the form of the iterative equations
\be 
\label{36}
 s^{(n+1)} = s_B \; \sqrt{n_0 + \sgm^{(n)} } \;  , \qquad
\sgm^{(n+1)} = \frac{s_B^2}{\pi^2} \; s^{(n+1)} \; ,
\ee
in which $n$ is an iteration number. Notice that these equations can be combined into
one iterative relation
\be
\label{37}
 \sgm^{(n+1)} = \frac{s_B^3}{\pi^2} \; \; \sqrt{n_0 + \sgm^{(n)} } \;  .
\ee

The Bogolubov approximation, with $s^{(0)} = s_B$ and $\sigma^{(0)} = 0$ can be 
accepted as the zero-order approximation for the iterative procedure. Then the first 
iteration gives
\be
\label{38}
 s^{(1)} = s_B\; \sqrt{n_0} \; , \qquad 
\sgm^{(1)} = \frac{s_B^3}{\pi^2}\; \sqrt{n_0} \;  .
\ee
This is equivalent to the approximation of Ref. [19] that, hence, can be considered
as the first iteration of the iterative procedure. To second order, we obtain
\be
\label{39}
s^{(2)} = s_B \left ( n_0 + \frac{s_B^3}{\pi^2}\; \sqrt{n_0} \right )^{1/2} \; , 
\qquad
\sgm^{(2)} = \frac{s_B^3}{\pi^2} \;
 \left ( n_0 + \frac{s_B^3}{\pi^2}\; \sqrt{n_0} \right )^{1/2} \;   .
\ee
In what follows, we shall use the second-order iteration for $\sigma$.

Summarizing the above consideration, we thus come to the system of equations
$$
n_0 = 1 - n_1 \; , \qquad n_1 = \frac{s^3}{3\pi^2} \; ,
$$
$$
s^2 = s_B^2 ( n_0 + \sgm ) \; ,
$$
\be
\label{40}      
\sgm =  \frac{s_B^3}{\pi^2} \; 
\left ( n_0 + \frac{s_B^3}{\pi^2}\; \sqrt{n_0} \right )^{1/2} \;  ,
\ee
self-consistently defining the condensate fraction $n_0$, fraction of uncondensed 
atoms $n_1$, sound velocity $s$, and the anomalous average $\sigma$. 

At small gas parameter $\gamma \ra 0$, we have 
$$
n_0 \simeq 1 - \; \frac{8}{3\sqrt{\pi}} \; \gm^{3/2} \; - \;
\frac{64}{3\pi} \; \gm^3 \; - \frac{256}{9\pi^{3/2}} \; \gm^{9/2} \; ,
$$
$$
n_1 \simeq \frac{8}{3\sqrt{\pi}} \; \gm^{3/2} \; + \;
\frac{64}{3\pi} \; \gm^3 \; + \; \frac{256}{9\pi^{3/2}} \; \gm^{9/2} \; ,
$$
$$
s \simeq \sqrt{4\pi\gm} \: + \; \frac{16}{3} \; \gm^2 \; - \; 
\frac{64}{9\sqrt{\pi}} \; \gm^{7/2} -\frac{4480}{27\pi} \; \gm^5 \; ,
$$
$$
\sgm \simeq \frac{8}{\sqrt{\pi}} \; \gm^{3/2} \; + \;
 \frac{64}{3\pi} \; \gm^3 \; - \; \frac{1408}{9\pi^{3/2}} \; \gm^{9/2} \;  .
$$
The first two terms in the expansion for the condensate fraction $n_0$ exactly
reproduce the Bogolubov behavior of $n_0$. We may notice that the anomalous average
is larger than the fraction of uncondensed atoms $n_1$, in particular
$$
\lim_{\gm\ra 0} \; \frac{\sgm}{n_1} = 3 \; .
$$
It is, therefore, would be mathematically incorrect to neglect $\sigma$ leaving the
three times smaller quantity $n_1$. The anomalous average is an important quantity,
without which the description would not be self-consistent and the system would be 
unstable.  

The behavior at large $\gamma$ can also be found from Eqs. (40). However, strictly 
speaking, considering $\gamma \gg 1$ is not applicable to a stable system, since  
it freezes at $\gamma = 0.653$, as follows from the Monte Carlo simulations [13].
But, keeping in mind a metastable situation, we can formally study large values of 
$\gamma \gg 1$, which leads to 
$$
n_0 \simeq 4 \times 10^{-5} \; \frac{1}{\gm^{13}} \; , \qquad
n_1 = 1 - n_0 \; ,
$$
$$
s \simeq  (3\pi^2)^{1/3} - \left ( \frac{\pi}{3} \right )^{2/3} \; n_0 \; ,
\qquad \sgm \simeq  \frac{(9\pi)^{1/3}}{4} \; \frac{1}{\gm}
\;  - \; n_0 \; .
$$
In the case of cold trapped atoms, although the scattering length can be made very 
large by means of Feshbach resonance, but such gases become unstable with respect 
to three-body recombination leading to significant particle loss and heating [36].      

We solve Eqs. (40) for arbitrary values of the gas parameter $\gamma$ and compare 
our results with the Monte Carlo simulations by Rossi and Salasnich [13]. The latter
confirm the earlier Monte Carlo calculations [7] and provide essentially more 
information for the larger values of the gas parameter. In Fig. 1, the behavior 
of the condensate fraction $n_0$ is shown, demonstrating good agreement with the 
Monte Carlo simulations [13] in the whole range of $\gamma$. The Bogolubov 
expression for the condensate fraction
$$
n_B = 1 \; - \; \frac{8}{3\sqrt{\pi}} \; \gm^{3/2}    
$$
is also shown. As is evident, $n_B$ gives a good approximation only for 
$\gamma < 0.1$ and for larger $\gamma$ is not applicable, deviating too strongly
from the numerical data. Figure 2 presents the fraction of uncondensed atoms $n_1$ 
and the anomalous average $\sigma$. As is seen, the latter is larger than the 
former in the whole range of the considered $\gamma$. In Fig. 3, the dimensionless 
sound velocity $s$ is compared with the Bogolubov sound velocity $s_B$. The former 
is larger than the latter, although their values are close to each other.   

There have been a number of attempts to measure the condensate fraction in superfluid
$^4$He with different experiments [37-41]. The estimated values of $n_0$ at zero 
temperature are in the range between $2\%$ and $10\%$. The most recent rather precise 
experiments [42-44] give the zero temperature value $n_0 = (7.25 \pm 0.75)\%$ at 
saturated vapor pressure and $n_0 = (2.8 \pm 0.2)\%$ at the pressure close to 
solidification. The latter value has also been confirmed by the diffusion Monte Carlo 
calculations [44]. The atoms of $^4$He at saturated vapor pressure can be well 
represented [6] by hard spheres of diameter $a_s = 2.203$ \AA, which corresponds to 
the gas parameter $\gamma = 0.59$. At this value, we get the condensate fraction  
about $3\%$.

\section{Ground-state energy}

The system ground-state energy is the internal energy at zero temperature
\be
\label{41}
 E = \lgl \hat H \rgl \qquad ( T = 0 ) \;  .
\ee
It is customary to express this energy in units of $\hbar^2/2ma_s^2$. In our 
notation, this gives the dimensionless ground-state energy
\be
\label{42}   
  E_0 \equiv 2ma_s^2 \; \frac{E}{N} \; .
\ee
Calculating the energy, we meet the divergent integral
$$
\int ( \ep_k - \om_k ) \; \frac{d\bk}{(2\pi)^3 } = 
\frac{16m^4}{15\pi^2} \; c_{eff}^5 \;  ,
$$
which is again regularized invoking dimensional regularization [31]. Then for
small gas parameters, we have
\be
\label{43}
 E_0 = 4\pi\gm^3 \left ( 1 + n_1^2 - 2n_1\sgm - \sgm^2 +
\frac{4s_{eff}^5}{15\pi^3\gm} \right ) \;  ,
\ee
which yields the asymptotic, as $\gamma \ra 0$, expansion 
\be
\label{44}
 E_0(\gm) \simeq  4\pi\gm^3 \left ( 1 + \frac{128}{15\sqrt{\pi}} \; \gm^{3/2} +
 \frac{128}{9\pi} \; \gm^3 \; - \; \frac{2048}{9\pi^{3/2}} \; \gm^{9/2} \right ) \; .
\ee
The first two terms in the right-hand side of Eq. (44) exactly coincide with the 
Lee-Huang-Yang formula [2-4]. The simplest way for extending this expression to the
larger values of the gas parameter is to use the extrapolation procedure based
on self-similar factor approximants [25]. To second order, we find
\be
\label{45}
E_0(\gm) = 4\pi \gm^3 ( 1 + 2.93379\gm^{3/2})^{1.64103} \;    .
\ee
This formula for $\gamma \ll 1$ reproduces exactly the Lee-Huang-Yang expression [2-4].
The behavior of the ground-state energy (45) is shown in Fig. 4, compared with the
Monte Carlo calculations by Rossi and Salasnich [13] and with the Lee-Huang-Yang 
perturbative expression
$$
E_{LHY} = 4\pi\gm^3 \left ( 1 + \frac{128}{15\sqrt{\pi}} \; \gm^{3/2}
\right ) \; .
$$
The agreement of our results with the Monte Carlo data [13] is good up to the 
values $\gamma \approx 0.6$. Let us recall that, actually, the system freezes [13] 
at $\gamma \approx 0.65$, so that to consider the gas parameters larger than 
the freezing value $0.65$ is not of much meaning. Let us emphasize that expression (45)
has been obtained without any fitting. The Lee-Huang-Yang values of $E_{LHY}$ give a
good approximation only for $\gamma < 0.4$, while our formula (45) yields the values
practically coinciding with the Monte Carlo Data [13] up to $\gamma = 0.6$.

\section{Conclusion}  

We have considered the ground state of a homogeneous Bose-condensed gas with a 
local pseudopotential imitating the hard-sphere interactions. The consideration 
is based on self-consistent mean-field approximation developed earlier by the 
authors. This approach allows one to extend the results obtained for small gas 
parameters to finite values of the latter. It is shown to be in good agreement 
with the accurate Monte Carlo results by Rossi and Salasnich [13] for all finite 
values of the gas parameter between zero and the point of freezing. The importance 
of using a correct expression for the anomalous average is emphasized. This 
explains why the previously used approximations could not provide sufficiently 
accurate behavior of the condensate fraction for finite gas parameters.

The main difference of the present paper from our previous publications is that 
here we have suggested an interative procedure for defining the anomalous average. 
The zeroth iteration of this procedure corresponds to the Bogolubov approximation, 
where the anomalous average is zero. This approximation is reasonable for small gas 
parameters $\gamma < 0.1$, but is not applicable for larger values of $\gamma$, as 
is evident from the comparison in the figures.  

The first iteration (38) corresponds to the expression we used in our earlier papers, 
which extends the applicability of the results to $\gamma \approx 0.4$. However, for 
the gas parameter larger than $0.4$, our previous results do not provide good 
approximation, as has been thoroughly analyzed in the paper by Rossi and Salasnich [13]. 

Now we have employed the second-order iteration (39), which have allowed us to 
essentially improve the results, making them very close to the numerical Monte Carlo 
data, as is demonstrated in the presented figures.     

Recently, we have demonstrated [45] that the self-consistent mean-field approach 
is the sole mean-field theory correctly describing Bose-Einstein condensation as 
a phase transition of second order for arbitrary values of the gas parameter. Now we 
have also proved that this approach provides quite accurate approximations for the 
condensate fraction and ground-state energy of the Bose system, being in good agreement 
with numerical Monte Carlo data [13].        

\vskip 5mm

{\bf Acknowledgments}

\vskip 3mm

The authors are grateful to L. Salasnich for useful discussions and for the 
detailed information on the data of his Monte Carlo calculations. Financial 
support from the Russian Foundation for Basic Research (grant $\# 14-02-00723$) 
is appreciated.

\newpage 

\begin{center}

{\Large{\bf Figure Captions } }

\end{center}

\vskip 1cm

Figure 1. Condensate fraction $n_0$ (solid line) as a function of the gas 
parameter $\gamma$, compared with the Monte Carlo results by Rossi and 
Salasnich [13], shown by dots, and with the Bogolubov approximation $n_B$
(dashed line). The latter is not applicable above $\gamma = 0.1$.    

\vskip 1cm
Figure 2. Fraction of uncondensed atoms $n_1$ (solid line) and anomalous 
average $\sigma$ (dashed-dotted line) as functions of the gas parameter $\gamma$.  
The anomalous average is everywhere larger than the $n_1$. 

\vskip 1cm
Figure 3. Sound velocity $s$ (solid line) in dimensionless units, compared 
with the Bogolubov sound velocity $s_B$ (dashed line), as functions of $\gamma$. 
The Bogolubov approximation essentially deviates from $s$ above $\gamma = 0.1$.

\vskip 1cm
Figure 4. Dimensionless ground-state energy $E_0$ (solid line) as a function 
of the gas parameter $\gamma$, compared with the Monte Carlo results by Rossi and 
Salasnich [13], shown by dots, and with the Lee-Huang-Yang expression $E_{LHY}$
(dashed line). The latter deviates from the numerical data after $\gamma = 0.4$.

\newpage

\begin{figure}[ht]
\centerline{\includegraphics[width=12cm]{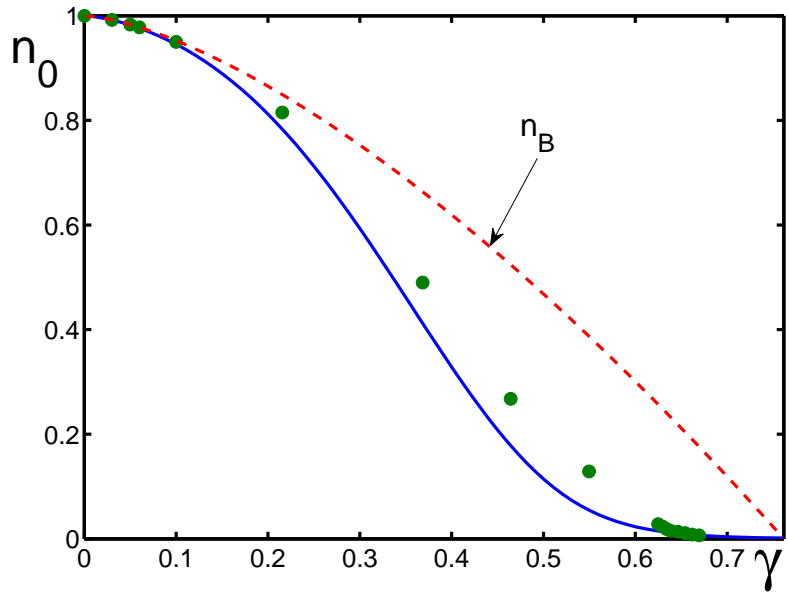} }
\caption{Condensate fraction $n_0$ (solid line) as a function of the gas 
parameter $\gamma$, compared with the Monte Carlo results by Rossi and 
Salasnich [13], shown by dots, and with the Bogolubov approximation $n_B$
(dashed line). The latter is not applicable above $\gamma = 0.1$.    
}
\label{fig:Fig.1}
\end{figure}

\newpage

\begin{figure}[ht]
\centerline{\includegraphics[width=12cm]{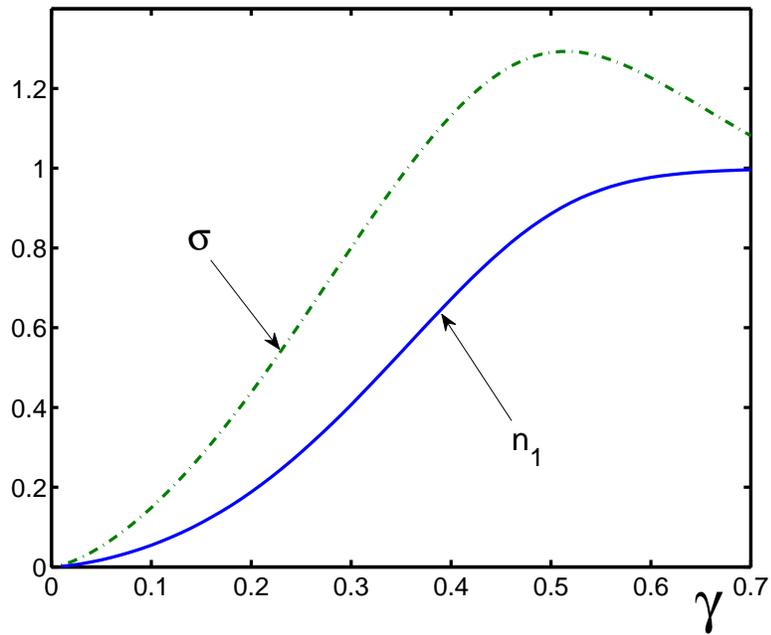} }
\caption{Fraction of uncondensed atoms $n_1$ (solid line) and anomalous 
average $\sigma$ (dashed-dotted line) as functions of the gas parameter $\gamma$.
The anomalous average is everywhere larger than the $n_1$.}
\label{fig:Fig.2}
\end{figure}

\newpage

\begin{figure}[ht]
\centerline{\includegraphics[width=10cm]{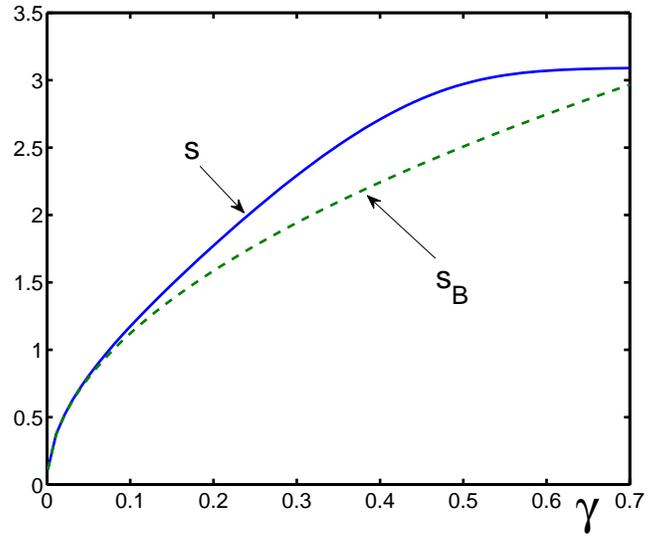} }
\caption{Sound velocity $s$ (solid line) in dimensionless units, compared with 
the Bogolubov sound velocity $s_B$ (dashed line), as functions of $\gamma$.
The Bogolubov approximation essentially deviates from $s$ above $\gamma = 0.1$.}
\label{fig:Fig.3}
\end{figure}

\newpage

\begin{figure}[ht]
\centerline{\includegraphics[width=10cm]{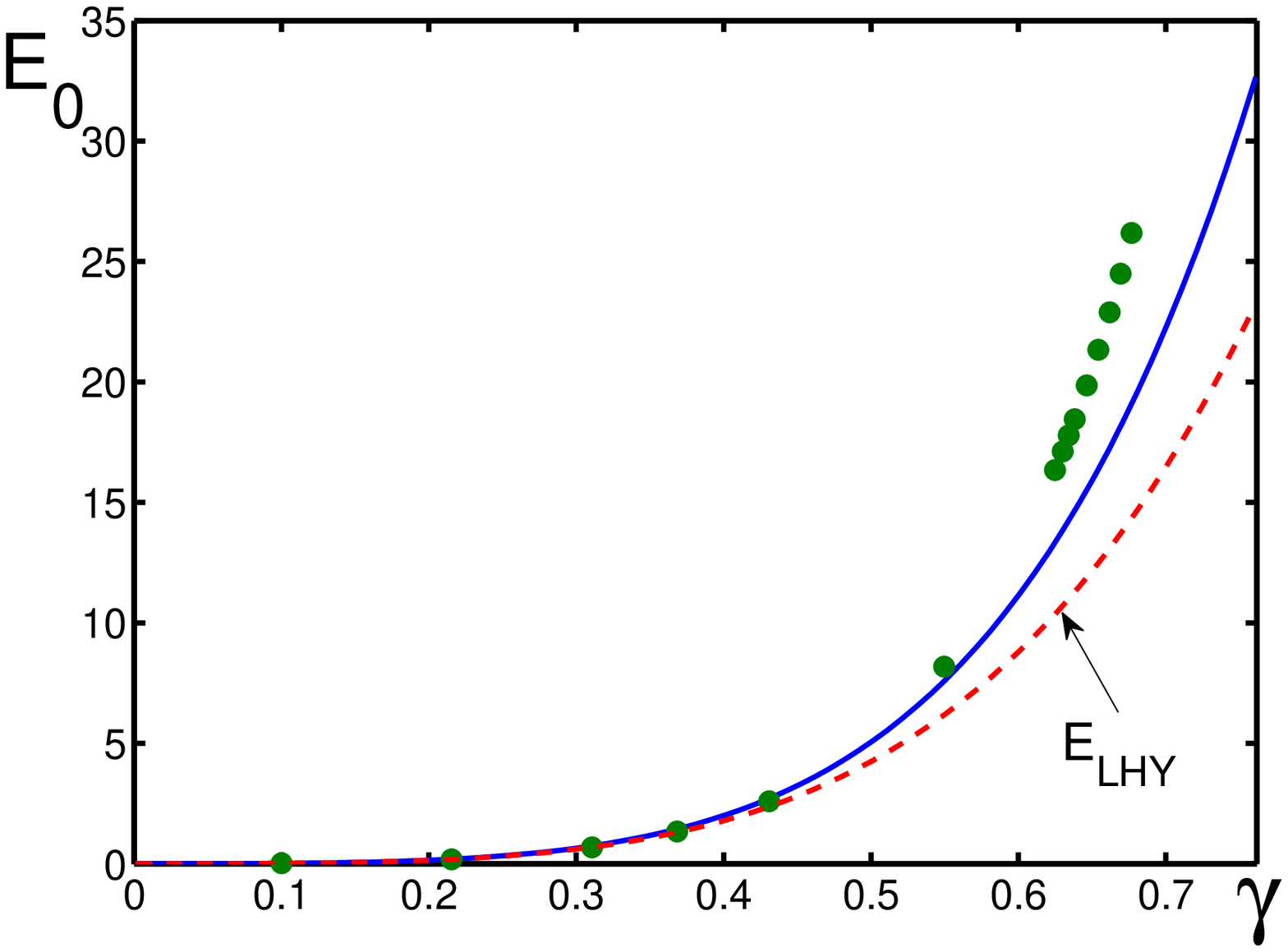} }
\caption{Dimensionless ground-state energy $E_0$ (solid line) as a function 
of the gas parameter $\gamma$, compared with the Monte Carlo results by Rossi and 
Salasnich [13], shown by dots, and with the Lee-Huang-Yang expression $E_{LHY}$
(dashed line). The latter deviates from the numerical data after $\gamma = 0.4$.}
\label{fig:Fig.4}
\end{figure}


\newpage


\begin{thebibliography}{99}
\bibitem{1}
N.N. Bogolubov, 
J. Phys. (Moscow) {\bf 11}, 23 (1947).  

\bibitem{2}
T.D. Lee and C.N. Yang, 
Phys. Rev. {\bf 105}, 1119 (1957).

\bibitem{3}
T.D. Lee, K. Huang, and C.N. Yang, 
Phys. Rev. {\bf 106}, 1135 (1957). 

\bibitem{4}
T.D. Lee and C.N. Yang, 
Phys. Rev. {\bf 112}, 1419 (1958).

\bibitem{5}
T.T. Wu,
Phys. Rev. {\bf 115}, 1390 (1959). 

\bibitem{6}
M.H. Kalos, D. Levesque, and L. Verlet,
Phys. Rev. A {\bf 9}, 2178 (1974). 

\bibitem{7}
S. Giorgini, J. Boronat, and J. Casulleras,
Phys. Rev. A {\bf 60}, 5129 (1999).  

\bibitem{8}
J.L. DuBois and H.R. Glyde,
Phys. Rev. A {\bf 63}, 023602 (2001). 

\bibitem{9}
J.L. DuBois and H.R. Glyde,
Phys. Rev. A {\bf 68}, 033602 (2003).

\bibitem{10}
W. Purwanto and S. Zhang,
Phys. Rev. A {\bf 72}, 053610 (2005).  

\bibitem{11}
K. Nho and D.P. Landau,
Phys. Rev. A {\bf 73}, 033606 (2006). 

\bibitem{12}
N. Navon, S. Piatecki, K. G\"{u}nter, B. Rem, T.C. Nguyen, F. Chevy, W. Krauth, 
and C. Salomon,
Phys. Rev. Lett. {\bf 107}, 135301 (2011).

\bibitem{13}
M. Rossi and L. Salasnich,
Phys. Rev. A {\bf 88}, 053617 (2013).  

\bibitem{14}
M. Rossi, L. Salasnich, P. Ancilotto, and F. Toigo,
Phys. Rev. A {\bf 89}, 041602 (2014).
 
\bibitem{15}
V.I. Yukalov and A.S. Shumovsky,
{\it Lectures on Phase Transitions} (World Scientific, Singapore, 1990). 

\bibitem{16}
J.M.D. Coey, 
{\it Magnetism and Magnetic Materials} (Cambridge University, Cambridge, 2010). 

\bibitem{17}
V.I. Yukalov,
Phys. Rev. E {\bf 72}, 066119 (2005).

\bibitem{18}
V.I. Yukalov,
Phys. Lett. A {\bf 359}, 712 (2006). 

\bibitem{19}
V.I. Yukalov and E.P. Yukalova,
Phys. Rev. A {\bf 74}, 063623 (2006). 

\bibitem{20}
V.I. Yukalov and E.P. Yukalova,
Phys. Rev. A {\bf 76}, 013602 (2007). 

\bibitem{21}
V.I. Yukalov,
Phys. Rep. {\bf 208}, 395 (1991). 

\bibitem{22}
V.I. Yukalov,
Ann. Phys. (N.Y.), {\bf 323}, 461 (2008). 

\bibitem{23}
K. Huang, C.N. Yang, and J.M. Luttinger,
Phys. Rev. {\bf 105}, 776 (1957).

\bibitem{24}
V.I. Yukalov and E.P. Yukalova,
Ann. Phys. (N.Y.) {\bf 277}, 219 (1999).

\bibitem{25}
V.I. Yukalov and E.P. Yukalova,
Phys. Lett. A {\bf 368}, 341 (2007). 

\bibitem{26}
E.H. Lieb, R. Seiringer, J.P. Solovej, and J. Yngvason,
{\it The Mathematics of the Bose Gas and Its Condensation} 
(Birkh\"{a}user, Basel, 2005). 

\bibitem{27}
V.I. Yukalov, 
Laser Phys. Lett. {\bf 4}, 632 (2007).

\bibitem{28}
N.N. Bogolubov,
{\it Lectures on Quantum Statistics} (Gordon and Breach, New York, 1967), Vol. 1.

\bibitem{29}
N.N. Bogolubov,
{\it Lectures on Quantum Statistics} (Gordon and Breach, New York, 1970), Vol. 2.

\bibitem{30}
V.I. Yukalov,
Laser Phys. {\bf 16}, 511 (2006).

\bibitem{31}
V.I. Yukalov,
Phys. Part. Nucl. {\bf 42}, 460 (2011). 

\bibitem{32}
V.I. Yukalov,
Laser Phys. {\bf 23}, 062001 (2013). 

\bibitem{33}
A. Rakhimov, C.K. Kim, S.H. Kim, and J.H. Yee,
Phys. Rev. A {\bf 77}, 033626 (2008). 

\bibitem{34}
J.O. Andersen, 
Rev. Mod. Phys. {\bf 76}, 599 (2004). 

\bibitem{35}
V.I. Yukalov and H. Kleinert,
Phys. Rev. A {\bf 73}, 063612 (2006). 

\bibitem{36}
R.J. Fletcher, A.I. Gaunt, N. Navon, R.P. Smith, and Z. Hadzibabic,
Phys. Rev. Lett. {\bf 111}, 125303 (2013). 

\bibitem{37}
H.A. Mook, R. Schern, and M.K. Wilkinson,
Phys. Rev. A {\bf 6}, 2268 (1972). 

\bibitem{38}
F.H. Wirth and R.B. Hallock,
Phys. Rev. B {\bf 35}, 89 (1987). 

\bibitem{39}
T.R. Sosnick, W.M. Snow, and P.E. Sokol, 
Phys. Rev. B {\bf 41}, 707 (1990). 

\bibitem{40}
R.T. Azuah, W.G. Stirling, H.R. Glyde, M. Boninsegni, P.E. Sokol, and S.M. Bennington, 
Phys. Rev. B {\bf 56}, 14620 (1997).

\bibitem{41}
A.S. Rinat and M.F. Taragin, 
J. Low. Temp. Phys. {\bf 123}, 139 (2001).

\bibitem{42}
H.R. Glyde, R.T. Azuah, and W.G. Stirling,
Phys. Rev. B {\bf 62}, 14337 (2000).

\bibitem{43}
H.R. Glyde, S.O. Diallo, R.T. Azuah, O. Kirichek, and J.W. Taylor,
Phys. Rev. B {\bf 84}, 184506 (2011).

\bibitem{44}
S.O. Diallo, R.T. Azuah, D.L. Abernathy. R. Rota, J. Boronat, and H.R. Glyde,
Phys. Rev. B {\bf 85}, 140505 (2012).
 
\bibitem{45}
V.I. Yukalov and E.P. Yukalova,
J. Phys. B {\bf 47}, 095302 (2014).
\end{thebibliography}
\end{document}